\documentclass{article}
\usepackage{graphicx}
\UseRawInputEncoding

\newcommand{\be}{\begin{equation}}
\newcommand{\ee}{\end{equation}}
\newcommand{\bea}{\begin{eqnarray}}
\newcommand{\eea}{\end{eqnarray}}
\newcommand{\beaa}{\begin{eqnarray*}}
\newcommand{\eeaa}{\end{eqnarray*}}

\newcommand{\BB}{{{\rm I} \kern -2pt \rlap {\rm B} \kern +8pt}}

\advance\textheight by \topskip
\makeatletter

\@addtoreset{equation}{section}
\def\section{\@startsection {section}{1}{\z@}{-3.5ex plus -1ex minus
 -.2ex}{2.3ex plus .2ex}{\large\bf\centering}}
\def\subsection{\@startsection{subsection}{2}{\z@}{-3.25ex plus -1ex minus -.2ex}{1.5ex plus .2ex}{\bf}}
\def\subsubsection{\@startsection{subsubsection}{3}{\z@}{-3.25ex plus -1ex minus -.2ex}{1.5ex plus .2ex}{\sl}}
\makeatother
\begin{document}

\baselineskip 18pt \parindent 12pt \parskip 10pt

\begin{titlepage}

\begin{center}
{\Large {\bf Dynamics of kink-soliton solutions for
$2+1$-dimensional sine-Gordon equation}}\\\vspace{1.5in}

{\large
U. Saleem\footnote{%
usman\_physics@yahoo.com}, H.  Sarfraz \footnote{%
hira phys@yahoo.com} and Y.~Hanif \footnote{%
yasir\_pmc@yahoo.com}}\vspace{0.15in}

{\small{\it Department of Physics, University of the Punjab,\\
Quaid-e-Azam Campus, Lahore-54590, Pakistan.}}
\end{center}

\vspace{1cm}
\begin{abstract}
In this paper we study the dynamics of explicit solutions of
$2+1$-dimensional ($2$D) sine-Gordon equation. The Darboux
transformation is applied to the associated linear eigenvalue
problem to construct nontrivial solutions of $2$D sine-Gordon
equation in terms of ratios of determinants. We obtained a
generalized expression for $N$-fold transformed dynamical variable
which enables us to calculate explicit expressions of nontrivial
solutions. In order to explore the dynamics of kink soliton
solutions explicit expressions one- and two-soliton solutions are
derived for particular column solutions. Different profiles of
kink-kink and, kink and anti-kink interactions are illustrated for a
different parameters and arbitrary functions. First-order bound
state solution is also displayed in our work.
\end{abstract}
\vspace{1cm}

\end{titlepage}

\section{Introduction}
\label{intro} A famous hyperbolic nonlinear partial differential
equation is the well-known sine-Gordon equation. The $1$D
sine-Gordon equation (SGE) reads \cite{Rajaraman}
\begin{equation}
\frac{\partial^2 s}{\partial Y^2}-\frac{\partial^2 s}{\partial
T^2}=\sin s, \label{SGE01}
\end{equation}%
where $s(Y, T)$ and $Y,T$ denote real scalar field  and, usual
spatial and time coordinates, respectively. The $1$D SGE is an
important integrable equation in the theory of nonlinear integrable
equations. Using light-cone coordinates
$y=\frac{1}{2}\left(Y+T\right), t=\frac{1}{2}\left(Y-T\right)$, the
$1$D SGE (\ref{SGE01}) can be written as
\begin{equation}
\frac{\partial^2 s}{\partial y\partial t}=\sin s. \label{SGE01LC}
\end{equation}%
The $1$D SGE (\ref{SGE01}) was first time appeared in 1862 during
the study of surfaces with constant negative curvature as the
Gauss-Codazzi equation for surfaces of the curvatures
\cite{Bour1862}. In 1875, a Swedish mathematician Albert Victor
B\"{a}cklund proposed a transformation which permits him to
construct a new surface with negative curvature from the known
\cite{Victor}. In fact, the B\"{a}cklund transformation is a
differential relation between two different solutions (or surfaces)
of the SGE. The B\"{a}cklund transformation allows to compute an
infinite many solutions of the SGE from simplest known solution.
Indeed, the B\"{a}cklund transformation obeys the theorem of
permutability which facilitates in the computation of higher-order
non-trivial solutions.

In 1938, Yakov Frenkel and Tatiana Kontorova proposed a mathematical
model
\begin{equation}
\frac{d^2 s_{n}}{d t^2}-C\left(s_{n+1}+s_{n-1}-2s_{n}\right)=\sin
s_{n},\label{SGE01dis}
\end{equation}
also known as Frenkel-Kontorova (FK) model which describe the
structure and dynamics of a crystal lattice near a dislocation
\cite{Frenkel}. Here $C$ is the elastic constant. Under the
continuum limit equation (\ref{SGE01dis}) reduces to $1$D SGE
(\ref{SGE01}).

There has been an increasing interest in the study of SGE due to
inherent rich mathematical structures. It is a relativistic $1$D
integrable equation with several applications in different fields
both mathematical and applied sciences. Ablowitz, Kaup, Newell and
Segure explored the integrability of the SGE (\ref{SGE01}) by means
of the inverse scattering transform method and obtained explicit
solutions \cite{Ablowitz1973}-\cite{Zakharov}. The solvability of
SGE was explored by using B\"{a}cklund transformation almost one
century earlier the discovery of the integrability of this equation
via inverse scattering transform method \cite{Victor}. Lamb in 1967
computed multi-soliton solutions of $1$D SGE (\ref{SGE01}) by
employing B\"{a}cklund transformation \cite{Lamb1967}. The SGE
(\ref{SGE01}) also shares various interesting properties, for
example, the existence of infinitely many conserved quantities,
multi-soliton solutions and Painlev\'{e} property. The $1$D SGE is
an important simplest integrable models like the Korteweg de Vries
and the nonlinear Schr\"{o}dinger (NLS) equations. The SGE has been
attracted a great deal of attention in different areas of applied
sciences such as propagation of ultra-short pulse in resonant laser
medium \cite{Lamb1967}, condensed matter physics
\cite{CMP1,CMP2,CMP3,CMP4,CMP5}, elementary particle physics
\cite{particle1,particle1a,particle2,particle3,particle4}, biology
\cite{BIO1,BIO2,BIO3,BIO4,BIO5} and fluid mechanics \cite{Gibbon}.

The $1$D SGE (\ref{SGE01LC}) can also be expressed as the
integrability condition of the following linear equations
\cite{Ablowitz1973}
\begin{eqnarray}
\frac{\partial \Phi}{\partial y}&=&P_{1}\Phi , \quad\quad\quad
\frac{\partial \Phi}{\partial t}=P_{2}\Phi , \label{linear2LC}
\end{eqnarray}%
where $\Phi (y, t;\lambda )=\left(
\begin{array}{cc}
\phi_{1} &
\phi_{2}%
\end{array}%
\right)^{T} $ is a column solution. The matrices $P_{1} (y, t;\lambda )$ and $P_{2} (y, t;\lambda )$ are given by%
\begin{equation}
P_{1} (y, t;\lambda ) =\left(
\begin{array}{cc}
\frac{\mbox{i}}{2}\frac{\partial s}{\partial y} & \lambda \\
\lambda & -\frac{\mbox{i}}{2}\frac{\partial s}{\partial y}%
\end{array}%
\right) ,  \quad \quad \quad \quad P_{2} (y, t;\lambda )
=\frac{1}{4\lambda }\left(
\begin{array}{cc}
0 & e^{\mbox{i}s} \\
e^{-\mbox{i}s} & 0%
\end{array}%
\right), \label{linear4LC}
\end{equation}%
where $\lambda$ is a real/complex-valued spectral parameter and
$\mbox{i}=\sqrt{-1}$. The integrability condition of the linear
system (\ref{linear2LC}), that is $\frac{\partial^2 \Psi}{\partial Y
\partial T} =\frac{\partial^2 \Psi}{\partial T \partial Y}
$ yields the zero-curvature condition $\frac{\partial P_{1}}{\partial T}-\frac{\partial P_{2}}{\partial Y}+[ P_{1},P%
_{2}] =0\Leftrightarrow [ \frac{\partial }{\partial
Y}-P_{1},\frac{\partial }{\partial T}-P_{2}] =0,$ here $[\quad ,
\quad ]$ represents matrix commutator.

Integrable equations are extensively used in almost all branches of
physics such as the field theory, theory of condensed matter
physics, plasma physics, optics, geophysics and nuclear and particle
physics, as well as in the other branches of science such as
chemistry, biology and communications. During the past few decades
integrable equations in $1$D have been studied from different view
points, for example, construction of multi-soliton solutions (by
using B\"{a}cklund-Darboux transformation, dressing method, Hirota's
direct method), conserved quantities, Hamiltonian structures,
Painlev\'{e} analysis \cite{book1,book2,book3}. Higher dimensional
integrable equations have attracted as great deal of attention in
recent years. The Nizhnik-Novikov-Veselov equation
\cite{Novikov1984} and Davey-Stewartson equation \cite{Stewartson}
are well known $2$D generalizations of the NLS and the KdV
equations, respectively. Recently, Wang \textit{et. al}
\cite{2DSGE1} have derived $2$D SGE from the AKNS system and
obtained nontrivial explicit solutions. The 2D integrable equations
have studied recent years \cite{2DSGE2,2DSGE3,2DSGE4,2DSGE5}. As we
have already mentioned above that the integrable $1$D SGE explains
different physical phenomenon, including the propagation of fluxons
in a junction between two superconductors (also known as Josephson
junction), the motion of coupled pendulum, dislocations in crystals
and dynamics of DNA (Deoxyribonucleic acid). Different integrability
aspects such as existence of infinitely many conserved quantities,
Painlev\'{e} property, multi-soliton solutions, Hirota
bilinearization of the $1$D SGE have been investigated. These
applications motivated us to explore the different integrable
features of $2$D SGE recently studied by Wang \textit{et. al}
\cite{2DSGE1}. The Darboux transformation in an elegant solution
generating technique among other solution generating techniques such
as inverse scattering transform method, Hirota direct method,
B\"{a}cklund transformation and dressing method
\cite{book1,book2,book3}. In this paper, we study the construction
of multi-soliton solutions of $2$D SGE by employing Darboux
transformation to the associated AKNS scheme. The expression of
$N$-fold dynamical variables are expressed in form of determinants.
We also obtain explicit expressions of one-, two-soliton solutions.
The dynamics of single-soliton (or kink/anti-kink) and different
type of interactions of two-soliton solutions such as kink and kink,
kink and anti-kink, and breather solutions have been investigated in
details for different choices of parameters and arbitrary functions.

The rest of this is organized as follows. In the section \ref{lax},
we express the $2$D SGE as an integrability condition of linear
system. In section $3$, we define a matrix Darboux transformation
and apply to the solution of the linear system and obtain
multi-soliton solutions in term of ratio of determinants. In section
$4$, we obtain explicit expressions of one-, two-kink and breather
soliton solutions of $2$D SGE. Last section is dedicated for
conclusion and open problems.

\section{Linear system and Darboux transformation for $2$-dimensional sine-Gordon equation}
\label{lax}

The $2$D SGE is given by%
\begin{equation}
\frac{\partial^2 s}{\partial x^2}-\frac{\partial^2 s}{\partial
x\partial y}-\frac{\partial^2 s}{\partial x\partial
t}+\frac{\partial^2 s}{\partial y\partial t}=\sin s, \label{SGE1}
\end{equation}%
where $s=s(x, y, t)$ is the field of the ultra-short optical pulses.

The $2$D SGE (\ref{SGE1}) is equivalent to the consistency condition
of the following matrix-valued linear system%
\begin{eqnarray}
\frac{\partial \Phi}{\partial y}&=&\frac{\partial \Phi}{\partial
x}+F_{1}\Phi , \quad\quad\quad \frac{\partial \Phi}{\partial
t}=\frac{\partial \Phi}{\partial x}+F_{2}\Phi .\label{linear2}
\end{eqnarray}%
The coefficient matrices $F_{1}(x, y, t;\lambda )$ and $F_{2}(x, y, t;\lambda )$ are defined as%
\begin{equation}
F_{1} = \left(
\begin{array}{cc}
\frac{\mbox{i}}{2}\left(\frac{\partial s}{\partial
y}-\frac{\partial s}{\partial x}\right) & \lambda \\
\lambda & -\frac{\mbox{i}}{2}\left(\frac{\partial s}{\partial
y}-\frac{\partial s}{\partial x}\right) %
\end{array}%
\right) ,  \quad \quad \quad \quad \quad F_{2} = \frac{1}{4\lambda
}\left(
\begin{array}{cc}
0 & e^{\mbox{i}s} \\
e^{-\mbox{i}s} & 0%
\end{array}%
\right), \label{linear4}
\end{equation}%
the consistency condition yields (\ref{SGE1}).

Under appropriate transformations the $2$D SGE (\ref{SGE1}) can also
be reexpressed in the form of $1$D SGE. If we define
$\xi=x+y+t,\eta=y,\tau=t$ and $s=S(\eta,\tau)$, the $2$D SGE
(\ref{SGE1}) reduces to
\begin{equation}
\frac{\partial^2 S}{\partial \eta\partial \tau}=\sin S,
\label{SGE1c}
\end{equation}%
which is the exactly the case of $1$D SGE.

The most important solution generating technique is the Darboux
transformation \cite{book1,book3}. Darboux transformation is a type
of gauge transformation such that associated linear eigenvalue
problem remains invariant. The one-fold Darboux transformation for
the system (\ref{linear2}) can be defined as
\begin{eqnarray}
\phi_1[1] &=&\lambda \phi_2 -\alpha^{(1)}\phi_1,  \label{sd-DT1a} \\
\phi_2[1] &=&\lambda \phi_1-\beta^{(1)}\phi_2, \label{sd-DT2a}
\end{eqnarray}
the unknown coefficients $\alpha^{(1)}$ and $\beta^{(1)} $ may be
obtained from the following conditions
\begin{eqnarray}
\left. \phi_1[1]\right|_{\lambda=\lambda_{1}, \phi_1=\phi_1^{(1)},\phi_2=\phi_2^{(1)}}&=&0,  \label{sd-DT1b} \\
\left. \phi_2[1]\right|_{\lambda=\lambda_{1},
\phi_1=\phi_1^{(1)},\phi_2 =\phi_2^{(1)}}&=&0. \label{sd-DT2b}
\end{eqnarray}
The above conditions allow us to write one-fold transformation
(\ref{sd-DT1a})-(\ref{sd-DT2a}) as
\begin{eqnarray}
\phi_1[1] &\equiv&\lambda \phi_2-\frac{\lambda _{1}\phi_2^{(1)}
}{\phi_1^{(1)}}%
\phi_1=\frac{\det\left(
                \begin{array}{cc}
 \lambda \phi_2 & \phi_1 \\
 \lambda _{1}\phi_2^{(1)}
& \phi_1^{(1)}
 \\
                \end{array}
              \right)
}{\phi_1^{(1)}},  \label{sd-DT1} \\
\phi_2[1] &\equiv&\lambda \phi_1-\frac{\lambda _{1}\phi_1^{(1)}}{\phi_2^{(1)}}%
\phi_2=\frac{\det\left(
                \begin{array}{cc}
 \lambda \phi_1 & \phi_2 \\
 \lambda _{1}\phi_1^{(1)}
& \phi_2^{(1)}
 \\
                \end{array}
              \right)
}{\phi_2^{(1)}},  \label{sd-DT2}
\end{eqnarray}
here $\phi_1^{(1)}$ and $\phi_2^{(1)}$ denote the particular
solutions to the linear system (\ref{linear2}). The linear system
(\ref{linear2}) is covariant under the action of Darboux
transformation (\ref{sd-DT1})-(\ref{sd-DT2}), that is,
\begin{eqnarray}
\frac{\partial \phi_{1}[1]}{\partial y} &=&\frac{\partial
\phi_{1}[1]}{\partial x}+\frac{\mbox{i}}{2}\left(\frac{\partial
s[1]}{\partial
y}-\frac{\partial s[1]}{\partial x}\right)\phi_1[1]+\lambda \phi_2[1],  \nonumber \\
\frac{\partial \phi_{2}[1]}{\partial y}&=&\frac{\partial
\phi_{2}[1]}{\partial x}+\lambda
\phi_1[1]-\frac{\mbox{i}}{2}\left(\frac{\partial s[1]}{\partial
y}-\frac{\partial s[1]}{\partial x}\right) \phi_2[1],  \label{Lax1 DT} \\
\frac{\partial \phi_{1}[1]}{\partial t}&=&\frac{\partial
\phi_{1}[1]}{\partial x}+\frac{1}{4\lambda}e^{\mbox{i}s[1]}\phi_2[1]\nonumber \\
\frac{\partial \phi_{2}[1]}{\partial t} &=&\frac{\partial
\phi_2[1]}{\partial x}+
\frac{1}{4\lambda}e^{-\mbox{i}s[1]}\phi_1[1]. \label{Lax2 DT}
\end{eqnarray}%
By substitution of scalar functions $\phi_1[1]$ and $\phi_2[1]$ from
(\ref{sd-DT1})-(\ref{sd-DT2}) in (\ref{Lax1 DT})-(\ref{Lax2 DT}), we
obtain
\begin{equation}
s\left[ 1\right] =s-2\mbox{i} \ln \left( \frac{\phi_2^{(1)}}{\phi_1
^{(1)}}\right). \label{sg16}
\end{equation}%
The two-fold Darboux transformation is defined as
\begin{eqnarray}
\phi_1[2] &\equiv&\lambda \phi_2[1]-\alpha^{(2)}[1]\phi_1[1]=\lambda^2\phi_1-\theta^{1}\lambda \phi_2-\theta^{0}\phi_1,  \label{Twosd-DT1a} \\
\phi_2[2] &\equiv&\lambda \phi_1[1]-\beta^{(2)}
[1]\phi_2[1]=\lambda^2\phi_2-\vartheta^{1}\lambda
\phi_1-\vartheta^{0}\phi_2, \label{Twosd-DT2a}
\end{eqnarray}
with $\theta^{1}=\beta^{(1)}+\alpha^{(2)}[1]$,
$\theta^{0}=-\alpha^{(2)}[1]\alpha^{(1)}$ and
$\vartheta^{1}=\alpha^{(1)}+\beta^{(2)}[1]$,
$\vartheta^{0}=-\beta^{(2)}[1]\beta^{(1)}$. The pair of unknown
coefficients $\theta^{0}$, $\theta^{1}$ and $\vartheta^{0}$,
$\vartheta^{1}$ can be determined from the following conditions
\begin{eqnarray}
\left. \phi_1[2]\right|_{\lambda=\lambda_{k}, \phi_1=\phi_1^{(k)},\phi_2=\phi_2^{(k)}}&=&0,  \label{Twosd-DT1b} \\
\left. \phi_2[2]\right|_{\lambda=\lambda_{k},
\phi_1=\phi_1^{(k)},\phi_2 =\phi_2^{(k)}}&=&0, \label{Twosd-DT2b}
\end{eqnarray}
for $k=1,2$. The above conditions reduce to following systems of
linear equations
\begin{eqnarray}
\left(
  \begin{array}{cc}
    \phi_1^{(1)} & \lambda_{1}\phi_2^{(1)} \\
    \phi_1^{(2)} & \lambda_{2}\phi_2^{(2)} \\
  \end{array}
\right)\left(
         \begin{array}{c}
           \theta^{0} \\
           \theta^{1} \\
         \end{array}
       \right)
&=&\left(
      \begin{array}{c}
        \lambda_{1}^2\phi_1^{(1)} \\
         \lambda_{2}^2\phi_1^{(2)} \\
      \end{array}
    \right)
,  \label{Twosd-DT1c} \\
\left(
  \begin{array}{cc}
    \phi_2^{(1)} & \lambda_{1}\phi_1^{(1)} \\
    \phi_2^{(2)} & \lambda_{2}\phi_1^{(2)} \\
  \end{array}
\right)\left(
         \begin{array}{c}
           \vartheta^{0} \\
           \vartheta^{1} \\
         \end{array}
       \right)
&=&\left(
      \begin{array}{c}
        \lambda_{1}^2\phi_2^{(1)} \\
         \lambda_{2}^2\phi_2^{(2)} \\
      \end{array}
    \right)
.\label{Twosd-DT2c}
\end{eqnarray}
On substitution of unknown coefficients in system of equations given
by (\ref{Twosd-DT1a})-(\ref{Twosd-DT2a}), we have
\begin{eqnarray}
\phi_1[2] &=&\frac{\det\left(
                \begin{array}{ccc}
 \lambda^{2}\phi_1 &\lambda \phi_2 &\phi_1\\
 \lambda_{1}^{2}\phi_1^{(1)} &\lambda_{1} \phi_2^{(1)} & \phi_1^{(1)} \\
 \lambda_{2}^{2}\phi_1^{(2)} &\lambda_{2} \phi_2^{(2)} &\phi_1^{(2)}
                \end{array}
              \right)
}{\det\left(
                \begin{array}{cc}
  \lambda_{1} \phi_2^{(1)} & \phi_1^{(1)} \\
 \lambda_{2} \phi_2^{(2)} & \phi_1^{(2)}
                \end{array}
              \right)},  \label{Twosd-DT1} \\
\phi_2[2] &=&\frac{\det\left(
                \begin{array}{ccc}
 \lambda^{2}\phi_2 & \lambda \phi_1 & \phi_2\\
 \lambda_{1}^{2}\phi_2^{(1)} & \lambda_{1} \phi_1^{(1)} & \phi_2^{(1)} \\
 \lambda_{2}^{2}\phi_2^{(2)} & \lambda_{2} \phi_1^{(2)} & \phi_2^{(2)}
                \end{array}
              \right)
}{\det\left(
                \begin{array}{cc}
  \lambda_{1} \phi_1^{(1)} & \phi_2^{(1)} \\
 \lambda_{2} \phi_1^{(2)} & \phi_2^{(2)}
                \end{array}
              \right)}.  \label{Twosd-DT2}
\end{eqnarray}

The two-fold transformed solutions given by
(\ref{Twosd-DT1})-(\ref{Twosd-DT2}) also satisfy the linear system
(\ref{linear2}), that is,
\begin{eqnarray}
\frac{\partial \phi_{1}[2]}{\partial y}&=&\frac{\partial
\phi_1[2]}{\partial x} +\frac{\mbox{i}}{2}\left(\frac{\partial s[2]}
{\partial y}-\frac{\partial s[2]}{\partial x}\right)\phi_1[2]+\lambda \phi_2[2],  \nonumber \\
\frac{\partial \phi_{2}[2]}{\partial y}&=&\frac{\partial
\phi_2[2]}{\partial x} +\lambda
\phi_1[2]-\frac{\mbox{i}}{2}\left(\frac{\partial s[2]}{\partial
y}-\frac{\partial s[2]}{\partial x}\right)\phi_2[2],  \label{Lax1 DT2} \\
\frac{\partial \phi_{1}[2]}{\partial
t}&=&\frac{\partial \phi_1[2]}{\partial x} + \frac{1}{4\lambda}e^{\mbox{i}s[2]}\phi_2[2],\nonumber \\
\frac{\partial \phi_{2}[2]}{\partial t}&=&\frac{\partial
\phi_2[2]}{\partial x}
+\frac{1}{4\lambda}e^{-\mbox{i}s[2]}\phi_1[2]. \label{Lax2 DT2}
\end{eqnarray}%
Using equations (\ref{Twosd-DT1})-(\ref{Twosd-DT2}) in equations
(\ref{Lax1 DT2})-(\ref{Lax2 DT2}), we obtain
\begin{equation}
s\left[2\right]=s+2\mbox{i}\ln\frac{\det\left(
                \begin{array}{cc}
  \lambda_{1} \phi_1^{(1)} & \phi_2^{(1)} \\
 \lambda_{2} \phi_1^{(2)} & \phi_2^{(2)}
                \end{array}
              \right)}{\det\left(
                \begin{array}{cc}
  \lambda_{1}\phi_2^{(1)} & \phi_1^{(1)}\\
\lambda_{2}\phi_2^{(2) } & \phi_1^{(2)}
                \end{array}
              \right)}. \label{Twosg16}
\end{equation}
Similarly, three-fold Darboux transformation is defined as
\begin{eqnarray}
\phi_1[3] &\equiv&\lambda \phi_2[2]-\alpha^{(3)}[2]\phi_1[2]=\lambda^3\phi_2-\theta^{2}\lambda^2 \phi_1-\theta^{1}\lambda \phi_2-\theta^{0} \phi_1,  \label{Threesd-DT1a} \\
\phi_2[2] &\equiv&\lambda \phi_1
[2]-\beta^{(3)}[2]\phi_2[2]=\lambda^3\phi_1-\vartheta^{2}\lambda^2
\phi_2-\vartheta^{1}\lambda \phi_1-\vartheta^{0} \phi_2.
\label{Threesd-DT2a}
\end{eqnarray}
The set of unknown coefficients
$\{\theta^{0},\theta^{1},\theta^{2}\vartheta^{0}, \vartheta^{1},
\vartheta^{2}\}$ can be determined from the following conditions
\begin{eqnarray}
\left. \phi_1[3]\right|_{\lambda=\lambda_{k}, \phi_1=\phi_1^{(k)},\phi_2=\phi_2^{(k)}}&=&0,  \label{Threesd-DT1b} \\
\left. \phi_2[3]\right|_{\lambda=\lambda_{k},
\phi_1=\phi_1^{(k)},\phi_2 =\phi_2^{(k)}}&=&0, \label{Threesd-DT2b}
\end{eqnarray}
for $k=1,2,3$. The above conditions can also be written as
\begin{eqnarray}
\left(
  \begin{array}{ccc}
    \phi_1^{(1)} & \lambda_{1}\phi_2^{(1)} &\lambda^{2}_{1}\phi_1^{(1)}\\
    \phi_1^{(2)} & \lambda_{2}\phi_2^{(2)} &\lambda^{2}_{2}\phi_1^{(2)}\\
     \phi_1^{(3)} & \lambda_{3}\phi_2^{(3)} &\lambda^{2}_{3}\phi_1^{(3)}
  \end{array}
\right)\left(
         \begin{array}{c}
           \theta^{0} \\
           \theta^{1} \\
           \theta^{2}
         \end{array}
       \right)
&=&\left(
      \begin{array}{c}
        \lambda_{1}^3\phi_2^{(1)} \\
         \lambda_{2}^3\phi_2^{(2)} \\
         \lambda_{3}^3\phi_2^{(3)}
      \end{array}
    \right)
,  \label{Threesd-DT1c} \\
\left(
  \begin{array}{ccc}
    \phi_2^{(1)} & \lambda_{1}\phi_1^{(1)} &\lambda^{2}_{1}\phi_2^{(1)}\\
    \phi_2^{(2)} & \lambda_{2}\phi_1^{(2)} &\lambda^{2}_{2}\phi_2^{(2)}\\
     \phi_2^{(3)} & \lambda_{3}\phi_1^{(3)} &\lambda^{2}_{3}\phi_2^{(3)}
  \end{array}
\right)\left(
         \begin{array}{c}
           \vartheta^{0} \\
           \vartheta^{1} \\
           \vartheta^{2}
         \end{array}
       \right)
&=&\left(
      \begin{array}{c}
        \lambda_{1}^3\phi_1^{(1)} \\
         \lambda_{2}^3\phi_1^{(2)} \\
         \lambda_{3}^3\phi_1^{(3)}
      \end{array}
    \right)
 .\label{Threesd-DT2c}
\end{eqnarray}
After substitution of unknown coefficients the expression of
three-fold transformation (\ref{Threesd-DT1a})-(\ref{Threesd-DT2a})
can also be expressed as ratio of determinants
\begin{eqnarray}
\phi_1[3] &=&\frac{\det\left(
                \begin{array}{cccc}
 \lambda^{3}\phi_2 & \lambda^{2} \phi_1 & \lambda \phi_2 & \phi_1\\
 \lambda_{1}^{3}\phi_2^{(1)} & \lambda_{1}^{2} \phi_1^{(1)} & \lambda_{1} \phi_2^{(1)} & \phi_1^{(1)}\\
 \lambda_{2}^{3}\phi_2^{(2)} & \lambda_{2}^{2} \phi_1^{(2)} & \lambda_{2} \phi_2^{(2)} & \phi_1^{(2)}\\
 \lambda_{3}^{3}\phi_2^{(3)} & \lambda_{3}^{2} \phi_1^{(3)} & \lambda_{3} \phi_2^{(3)} & \phi_1^{(3)}
                \end{array}
              \right)
}{\det\left(
                \begin{array}{ccc}
   \lambda_{1}^{2} \phi_1^{(1)} & \lambda_{1} \phi_2^{(1)} & \phi_1^{(1)}\\
 \lambda_{2}^{2} \phi_1^{(2)} & \lambda_{2} \phi_2^{(2)} & \phi_1^{(2)}\\
 \lambda_{3}^{2} \phi_1^{(3)} & \lambda_{3} \phi_2^{(3)} & \phi_1^{(3)}
                \end{array}
              \right)},  \label{Threesd-DT1} \\
\phi_2[3] &=&\frac{\det\left(
                \begin{array}{cccc}
  \lambda^{3}\phi_1 & \lambda^{2} \phi_2 & \lambda \phi_1 & \phi_2\\
 \lambda_{1}^{3}\phi_1^{(1)} & \lambda_{1}^{2} \phi_2^{(1)} & \lambda_{1} \phi_1^{(1)} & \phi_2^{(1)}\\
 \lambda_{2}^{3}\phi_1^{(2)} & \lambda_{2}^{2} \phi_2^{(2)} & \lambda_{2} \phi_1^{(2)} & \phi_2^{(2)}\\
 \lambda_{3}^{3}\phi_1^{(3)} & \lambda_{3}^{2} \phi_2^{(3)} & \lambda_{3} \phi_1^{(3)} & \phi_2^{(3)}
                \end{array}
              \right)
}{\det\left(
                \begin{array}{ccc}
  \lambda_{1}^{2} \phi_2^{(1)} & \lambda_{1} \phi_1^{(1)} & \phi_2^{(1)}\\
 \lambda_{2}^{2} \phi_2^{(2)} & \lambda_{2} \phi_1^{(2)} & \phi_2^{(2)}\\
 \lambda_{3}^{2} \phi_2^{(3)} & \lambda_{3} \phi_1^{(3)} & \phi_2^{(3)}
                \end{array}
              \right)}.  \label{Threesd-DT2}
\end{eqnarray}
Three-fold transformed solutions $\phi_1[3]$ and $\phi_2[3]$ satisfy
the linear system (\ref{linear2}) such as
\begin{eqnarray}
\frac{\partial \phi_{1}[3]}{\partial y}&=&\frac{\partial
\phi_1[3]}{\partial x} +\frac{\mbox{i}}{2}\left(\frac{\partial s[3]}
{\partial y}-\frac{\partial s[3]}{\partial x}\right)\phi_1[3]+\lambda \phi_2[3],  \nonumber \\
\frac{\partial \phi_{2}[3]}{\partial y}&=&\frac{\partial
\phi_2[3]}{\partial x} +\lambda
\phi_1[3]-\frac{\mbox{i}}{2}\left(\frac{\partial s[3]}{\partial
y}-\frac{\partial s[3]}{\partial x}\right)\phi_2[3],  \label{Lax1 DT3} \\
\frac{\partial \phi_{1}[3]}{\partial
t}&=&\frac{\partial \phi_1[3]}{\partial x} + \frac{1}{4\lambda}e^{\mbox{i}s[3]}\phi_2[3],\nonumber \\
\frac{\partial \phi_{2}[3]}{\partial t}&=&\frac{\partial
\phi_2[3]}{\partial x}
+\frac{1}{4\lambda}e^{-\mbox{i}s[3]}\phi_1[3]. \label{Lax2 DT3}
\end{eqnarray}%
Using equations (\ref{Threesd-DT1})-(\ref{Threesd-DT2}) in
(\ref{Lax1 DT3})-(\ref{Lax2 DT3}), we obtain
\begin{equation}
s\left[3\right]=s-2\mbox{i}\ln\frac{\det\left(
                \begin{array}{ccc}
  \lambda_{1}^{2} \phi_2^{(1)} & \lambda_{1} \phi_1^{(1)} & \phi_2^{(1)} \\
 \lambda_{2}^{2} \phi_2^{(2)} & \lambda_{2} \phi_1^{(2)}
& \phi_2^{(2)}\\
 \lambda_{3}^{2} \phi_2^{(3)} & \lambda_{3} \phi_1^{(3)} & \phi_2^{(3)}
                \end{array}
              \right)}{\det\left(
                \begin{array}{ccc}
    \lambda_{1}^{2} \phi_1^{(1)} & \lambda_{1} \phi_2^{(1)} & \phi_1^{(1)} \\
 \lambda_{2}^{2} \phi_1^{(2)} & \lambda_{2} \phi_2^{(2)} & \phi_1^{(2)}\\
 \lambda_{3}^{2} \phi_1^{(3)} & \lambda_{3} \phi_2^{(3)} & \phi_1^{(3)}
                \end{array}
              \right)}. \label{Threesg16}
\end{equation}%

In what follows, we would like to generalize our results for
$N$-times iteration of Darboux transformation. For this purpose,
first take $N=2K$, the Darboux transformation can be factorize as
\begin{eqnarray}
\phi_1[2K] &=&\lambda^{2K}\phi_1-\theta^{2K-1}\lambda^{2K-1} \phi_2-\dots -\theta^{1}\lambda \phi_2-\theta^{0}\phi_1,  \label{2K1} \\
\phi_2[2K]
&=&\lambda^{2K}\phi_2-\vartheta^{2K-1}\lambda^{2K-1}\phi_1-\dots-\vartheta^{1}\lambda
\phi_1-\vartheta^{0}\phi_2,\label{2K2}
\end{eqnarray}
where the unknown coefficients can be computed from the following
conditions
\begin{eqnarray}
\left. \phi_1[2K]\right|_{\lambda=\lambda_{k}, \phi_1=\phi_1^{(k)},\phi_2=\phi_2^{(k)}}&=&0,  \label{Two2Ksd-DT1b} \\
\left. \phi_2[2K]\right|_{\lambda=\lambda_{k},
\phi_1=\phi_1^{(k)},\phi_2 =\phi_2^{(k)}}&=&0, \label{Two2Ksd-DT2b}
\end{eqnarray}
for $k=1,2,\dots, 2K$. The above conditions reduce to following
systems of linear equations \scriptsize
\begin{eqnarray}
\left(
  \begin{array}{ccccc}
    \phi_1^{(1)} & \lambda_{1}\phi_2^{(1)} & \dots &\lambda_{1}^{2K-2}\phi_1^{(1)} & \lambda_{1}^{2K-1}\phi_2^{(1)} \\
    \phi_1^{(2)} & \lambda_{2} \phi_2^{(2)} & \dots &\lambda_{2}^{2K-2}\phi_1^{(2)} & \lambda_{2}^{2K-1}\phi_2^{(2)} \\
    \vdots&\vdots&\ddots&\vdots&\vdots\\
    \phi_1^{(2K-1)} &  \lambda_{2K-1}\phi_2^{(2K-1)} & \dots &\lambda_{2K-1}^{2K-2}\phi_1^{(2K-1)} & \lambda_{2K-1}^{2K-1}\phi_2^{(2K-1)} \\
    \phi_1^{(2K)} &  \lambda_{2K}\phi_2^{(2K)} & \dots &\lambda_{2K}^{2K-2}\phi_1^{(2K)} & \lambda_{2K}^{2K-1}\phi_2^{(2K)}
    \end{array}
\right)\left(
         \begin{array}{c}
           \theta^{0} \\
           \theta^{1} \\
           \vdots\\
           \theta^{2K-2} \\
           \theta^{2K-1} \\
         \end{array}
       \right)
&=&\left(
      \begin{array}{c}
        \lambda_{1}^{2K}\phi_1^{(1)} \\
         \lambda_{2}^{2K}\phi_1^{(2)} \\
           \vdots\\
           \lambda_{2K-1}^{2K}\phi_1^{(2K-1)} \\
         \lambda_{2K}^{2K}\phi_1^{(2K)} \\
      \end{array}
    \right)
, \nonumber \\ \label{Two2Ksd-DT1c} \\
\left(
  \begin{array}{ccccc}
    \phi_2^{(1)} & \lambda_{1}\phi_1^{(1)}&\dots &\lambda_{1}^{2K-2}\phi_2^{(1)} & \lambda_{1}^{2K-1}\phi_1^{(1)} \\
    \phi_2^{(2)} & \lambda_{2}\phi_1^{(2)}&\dots &\lambda_{2}^{2K-2}\phi_2^{(2)} & \lambda_{2}^{2K-1}\phi_1^{(2)} \\
    \vdots&\vdots&\ddots&\vdots&\vdots\\
    \phi_2^{(2K-1)} & \lambda_{2K-1}\phi_1^{(2K-1)}&\dots &\lambda_{2K-1}^{2K-2}\phi_2^{(2K-1)} & \lambda_{2K-1}^{2K-1}\phi_1^{(2K-1)} \\
    \phi_2^{(2K)} & \lambda_{2K}\phi_1^{(2K)}&\dots &\lambda_{2K}^{2K-2}\phi_2^{(2K)} & \lambda_{2K}^{2K-1}\phi_1^{(2K)}
    \end{array}
\right)\left(
         \begin{array}{c}
           \vartheta^{0} \\
           \vartheta^{1} \\
           \vdots\\
           \vartheta^{2K-2} \\
           \vartheta^{2K-1} \\
         \end{array}
       \right)
&=&\left(
      \begin{array}{c}
        \lambda_{1}^{2K}\phi_2^{(1)} \\
         \lambda_{2}^{2K}\phi_2^{(2)} \\
           \vdots\\
           \lambda_{2K-1}^{2K}\phi_2^{(2K-1)} \\
         \lambda_{2K}^{2K}\phi_2^{(2K)} \\
      \end{array}
    \right)
, \nonumber\\.\label{Two2Ksd-DT2c}
\end{eqnarray}
\normalsize
 Using the values of unknown coefficients the $2K$-fold transformation becomes (\ref{2K1})-(\ref{2K2}) as
\begin{eqnarray}
\phi_1[2K] &=&\frac{\det\left(
                \begin{array}{ccccc}
 \lambda^{2K}\phi_1 &\lambda^{2K-1} \phi_2&\dots& \lambda \phi_2&\phi_1 \\
 \lambda_{1}^{2K}\phi_1^{(1)} &\lambda_{1}^{2K-1} \phi_2^{(1)}&\dots& \lambda_{1}\phi_2^{(1)}&\phi_1^{(1)}\\
 \vdots&\vdots&\ddots&\vdots&\vdots\\
 \lambda_{2K-1}^{2K}\phi_1^{(2K-1)} &\lambda_{2K-1}^{2K-1} \phi_2^{(2K-1)}&\dots&
 \lambda_{2K-1}\phi_2^{(2K-1)}&\phi_1^{(2K-1)}\\
 \lambda_{2K}^{2K}\phi_1^{(2K)} &\lambda_{2K}^{2K-1} \phi_2^{(2K)}&\dots& \lambda_{2K}\phi_2^{(2K)}&\phi_1^{(2K)}
                \end{array}
              \right)
}{\det\left(
                \begin{array}{cccccccc}
 \lambda_{1}^{2K-1} \phi_2^{(1)}& \lambda_{1}^{2K-2} \phi_1^{(1)}&\dots& \lambda_{1}\phi_2^{(1)}&\phi_1^{(1)}\\
 \lambda_{2}^{2K-1} \phi_2^{(2)}&\lambda_{2}^{2K-2} \phi_1^{(2)}&\dots& \lambda_{2}\phi_2^{(2)}&\phi_1^{(2)}\\
 \vdots&\vdots&\ddots&\vdots&\vdots\\
 \lambda_{2K-1}^{2K-1} \phi_2^{(2K-1)}&\lambda_{2K-1}^{2K-2} \phi_1^{(2K-1)}&\dots&
 \lambda_{2K-1}\phi_2^{(2K-1)}&\phi_1^{(2K-1)}\\
 \lambda_{2K}^{2K-1} \phi_2^{(2K)}&\lambda_{2K}^{2K-2} \phi_1^{(2K)}&\dots& \lambda_{2K}\phi_2^{(2K)}&\phi_1^{(2K)}
                \end{array}
              \right)},  \label{TwoKsd-DT1} \\
\phi_2[2K] &=&\frac{\det\left(
                \begin{array}{ccccc}
 \lambda^{2K}\phi_2 &\lambda^{2K-1} \phi_1&\dots& \lambda \phi_1&\phi_2\\
 \lambda_{1}^{2K}\phi_2^{(1)} &\lambda_{1}^{2K-1} \phi_1^{(1)}&\dots& \lambda_{1}\phi_1^{(1)}&\phi_2^{(1)}\\
 \vdots&\vdots&\ddots&\vdots&\vdots\\
 \lambda_{2K-1}^{2K}\phi_2^{(2K-1)} &\lambda_{2K-1}^{2K-1} \phi_1^{(2K-1)}&\dots&
 \lambda_{2K-1}\phi_1^{(2K-1)}&\phi_2^{(2K-1)}\\
 \lambda_{2K}^{2K}\phi_2^{(2K)} &\lambda_{2K}^{2K-1} \phi_1^{(2K)}&\dots& \lambda_{2K}\phi_1^{(2K)}&\phi_2^{(2K)}
                \end{array}
              \right)
}{\det\left(
                \begin{array}{cccccccc}
 \lambda_{1}^{2K-1} \phi_1^{(1)}& \lambda_{1}^{2K-2} \phi_2^{(1)}&\dots& \lambda_{1}\phi_1^{(1)}&\phi_2^{(1)}\\
 \lambda_{2}^{2K-1} \phi_1^{(2)}&\lambda_{2}^{2K-2} \phi_2^{(2)}&\dots& \lambda_{2}\phi_1^{(2)}&\phi_2^{(2)}\\
 \vdots&\vdots&\ddots&\vdots&\vdots\\
 \lambda_{2K-1}^{2K-1} \phi_1^{(2K-1)}&\lambda_{2K-1}^{2K-2} \phi_2^{(2K-1)}&\dots&
 \lambda_{2K-1}\phi_1^{(2K-1)}&\phi_2^{(2K-1)}\\
 \lambda_{2K}^{2K-1} \phi_1^{(2K)}&\lambda_{2K}^{2K-2} \phi_2^{(2K)}&\dots& \lambda_{2K}\phi_1^{(2K)}&\phi_2^{(2K)}
                \end{array}
              \right)}.  \label{TwoKsd-DT2}
\end{eqnarray}
The $2K$-fold transformed solutions given by
(\ref{TwoKsd-DT1})-(\ref{TwoKsd-DT2}) satisfy the linear system
(\ref{linear2})
\begin{eqnarray}
\frac{\partial \phi_1[2K] }{\partial y} &=&\frac{\partial \phi_1[2K]
}{\partial x}+\frac{\mbox{i}}{2}\left(\frac{\partial s[2K]}{\partial
y}-\frac{\partial s[2K]}{\partial x}\right)\phi_1[2K]+\lambda \phi_2[2K],  \nonumber \\
\frac{\partial \phi_2[2K]}{\partial y} &=&\frac{\partial
\phi_2[2K]}{\partial x}+\lambda
\phi_1[2K]-\frac{\mbox{i}}{2}\left(\frac{\partial s[2K]}{\partial
y}-\frac{\partial s[2K]}{\partial x}\right)\phi_2[2K],  \label{TwosdK-Lax1 DT} \\
\frac{\partial \phi_1[2K] }{\partial t} &=&\frac{\partial \phi_1[2K]
}{\partial x}+ \frac{1}{4\lambda}e^{\mbox{i}s[2K]}\phi_2[2K],  \nonumber \\
\frac{\partial \phi_2[2K]}{\partial t} &=&\frac{\partial
\phi_2[2K]}{\partial
x}+\frac{1}{4\lambda}e^{-\mbox{i}s[2K]}\phi_1[2K].
\label{TwosdK-Lax2 DT}
\end{eqnarray}%

From the covariance of the Darboux transformation, one can obtain
$2K$-fold transformed dynamical variable as
\begin{eqnarray}
s\left[2K\right] &=&s+2\mbox{i}\ln\frac{\det\left(
                \begin{array}{cccccccc}
 \lambda_{1}^{2K-1} \phi_1^{(1)}& \lambda_{1}^{2K-2} \phi_2^{(1)}&\dots& \lambda_{1}\phi_1^{(1)}&\phi_2^{(1)}\\
 \lambda_{2}^{2K-1} \phi_1^{(2)}&\lambda_{2}^{2K-2} \phi_2^{(2)}&\dots& \lambda_{2}\phi_1^{(2)}&\phi_2^{(2)}\\
 \vdots&\vdots&\ddots&\vdots&\vdots\\
 \lambda_{2K-1}^{2K-1} \phi_1^{(2K-1)}&\lambda_{2K-1}^{2K-2} \phi_2^{(2K-1)}&\dots&
 \lambda_{2K-1}\phi_1^{(2K-1)}&\phi_2^{(2K-1)}\\
 \lambda_{2K}^{2K-1} \phi_1^{(2K)}&\lambda_{2K}^{2K-2} \phi_2^{(2K)}&\dots& \lambda_{2K}\phi_1^{(2K)}&\phi_2^{(2K)}
                \end{array}
              \right)}{\det\left(
                \begin{array}{cccccccc}
 \lambda_{1}^{2K-1} \phi_2^{(1)}& \lambda_{1}^{2K-2} \phi_1^{(1)}&\dots& \lambda_{1}\phi_2^{(1)}&\phi_1^{(1)}\\
 \lambda_{2}^{2K-1} \phi_2^{(2)}&\lambda_{2}^{2K-2} \phi_1^{(2)}&\dots& \lambda_{2}\phi_2^{(2)}&\phi_1^{(2)}\\
 \vdots&\vdots&\ddots&\vdots&\vdots\\
 \lambda_{2K-1}^{2K-1} \phi_2^{(2K-1)}&\lambda_{2K-1}^{2K-2} \phi_1^{(2K-1)}&\dots&
 \lambda_{2K-1}\phi_2^{(2K-1)}&\phi_1^{(2K-1)}\\
 \lambda_{2K}^{2K-1} \phi_2^{(2K)}&\lambda_{2K}^{2K-2} \phi_1^{(2K)}&\dots& \lambda_{2K}\phi_2^{(2K)}&\phi_1^{(2K)}
                \end{array}
              \right)}.\nonumber\\ \label{2Kfold-sg16}
\end{eqnarray}%

Similarly, for $N=2K+1$, we have
\begin{eqnarray}
\phi_1[2K+1] &=&\lambda^{2K+1}\phi_2-\theta^{2K}\lambda^{2K} \phi_1-\dots -\theta^{1}\lambda \phi_2-\theta^{0}\phi_1,  \label{2K3} \\
\phi_2[2K+1]
&=&\lambda^{2K+1}\phi_1-\vartheta^{2K}\lambda^{2K}\phi_2-\dots-\vartheta^{1}\lambda
\phi_1-\vartheta^{0}\phi_2,\label{2K4}
\end{eqnarray}
with
\begin{eqnarray}
\left. \phi_1[2K+1]\right|_{\lambda=\lambda_{k}, \phi_1=\phi_1^{(k)},\phi_2=\phi_2^{(k)}}&=&0,  \label{TwoKsd-DT1b} \\
\left. \phi_2[2K+1]\right|_{\lambda=\lambda_{k},
\phi_1=\phi_1^{(k)},\phi_2 =\phi_2^{(k)}}&=&0, \label{TwoKsd-DT2b}
\end{eqnarray}
for $k=1,2,\dots,2K+1$. The above conditions can also be expressed
in matrix notation as  \scriptsize
\begin{eqnarray}
\left(
  \begin{array}{ccccc}
    \phi_1^{(1)} & \lambda_{1}\phi_2^{(1)}&\dots &\lambda_{1}^{2K-1}\phi_2^{(1)} & \lambda_{1}^{2K}\phi_1^{(1)} \\
    \phi_1^{(2)} & \lambda_{2}\phi_2^{(2)}&\dots &\lambda_{2}^{2K-1}\phi_2^{(2)} & \lambda_{2}^{2K}\phi_1^{(2)} \\
    \vdots&\vdots&\ddots&\vdots&\vdots\\
    \phi_1^{(2K)} &\lambda_{2K} \phi_2^{(2K)}&\dots &\lambda_{2K}^{2K-1}\phi_2^{(2K-1)} & \lambda_{2K}^{2K}\phi_1^{(2K)} \\
    \phi_1^{(2K+1)} & \lambda_{2K+1}\phi_2^{(2K+1)}&\dots &\lambda_{2K+1}^{2K-1}\phi_2^{(2K+1)} & \lambda_{2K+1}^{2K}\phi_1^{(2K+1)}
    \end{array}
\right)\left(
         \begin{array}{c}
           \theta^{0} \\
           \theta^{1} \\
           \vdots\\
           \theta^{2K-1} \\
           \theta^{2K} \\
         \end{array}
       \right)
&=&\left(
      \begin{array}{c}
        \lambda_{1}^{2K+1}\phi_2^{(1)} \\
         \lambda_{2}^{2K+1}\phi_2^{(2)} \\
           \vdots\\
           \lambda_{2K}^{2K+1}\phi_2^{(2K)} \\
         \lambda_{2K+1}^{2K+1}\phi_2^{(2K+1)} \\
      \end{array}
    \right)
, \nonumber\\ \label{TwoKsd-DT1c} \\
\left(
  \begin{array}{ccccc}
    \phi_2^{(1)} & \lambda_{1}\phi_1^{(1)}&\dots &\lambda_{1}^{2K-1}\phi_1^{(1)} & \lambda_{1}^{2K}\phi_2^{(1)} \\
    \phi_2^{(2)} & \lambda_{2}\phi_1^{(2)}&\dots &\lambda_{2}^{2K-1}\phi_1^{(2)} & \lambda_{2}^{2K}\phi_2^{(2)} \\
    \vdots&\vdots&\ddots&\vdots&\vdots\\
    \phi_2^{(2K)} &\lambda_{2K} \phi_1^{(2K)}&\dots &\lambda_{2K}^{2K-1}\phi_1^{(2K)} & \lambda_{2K}^{2K}\phi_2^{(2K)} \\
    \phi_2^{(2K+1)} & \lambda_{2K+1}\phi_1^{(2K+1)}&\dots &\lambda_{2K+1}^{2K-1}\phi_1^{(2K+1)} & \lambda_{2K+1}^{2K}\phi_2^{(2K+1)}
    \end{array}
\right)\left(
         \begin{array}{c}
           \vartheta^{0} \\
           \vartheta^{1} \\
           \vdots\\
           \vartheta^{2K-1} \\
           \vartheta^{2K} \\
         \end{array}
       \right)
&=&\left(
      \begin{array}{c}
        \lambda_{1}^{2K+1}\phi_1^{(1)} \\
         \lambda_{2}^{2K+1}\phi_1^{(2)} \\
           \vdots\\
           \lambda_{2K}^{2K+1}\phi_1^{(2K)} \\
         \lambda_{2K+1}^{2K+1}\phi_1^{(2K+1)} \\
      \end{array}
    \right)
, \nonumber\\.\label{TwoKsd-DT2c}
\end{eqnarray}
\normalsize Using the values of unknown coefficients the
$(2K+1)$-fold transformation (\ref{2K3})-(\ref{2K4}) can be
expressed as
\begin{eqnarray}
\phi_1[2K+1] &=&\frac{\det\left(
                \begin{array}{ccccc}
 \lambda^{2K+1}\phi_2 &\lambda^{2K} \phi_1 &\dots& \lambda \phi_2&\phi_1\\
 \lambda_{1}^{2K+1}\phi_2^{(1)} &\lambda_{1}^{2K} \phi_1^{(1)}&\dots& \lambda_{1}\phi_2^{(1)}&\phi_1^{(1)}\\
 \vdots&\vdots&\ddots&\vdots&\vdots\\
 \lambda_{2K}^{2K+1}\phi_2^{(2K)} &\lambda_{2K}^{2K} \phi_1^{(2K)}&\dots&
 \lambda_{2K}\phi_2^{(2K)}&\phi_1^{(2K)}\\
 \lambda_{2K+1}^{2K+1}\phi_2^{(2K+1)} &\lambda_{2K+1}^{2K} \phi_1^{(2K+1)}&\dots& \lambda_{2K+1}\phi_2^{(2K+1)}&\phi_1^{(2K+1)}
                \end{array}
              \right)
}{\det\left(
                \begin{array}{ccccc}
\lambda_{1}^{2K} \phi_1^{(1)}&\lambda_{1}^{2K-1} \phi_2^{(1)}&\dots& \lambda_{1}\phi_2^{(1)}&\phi_1^{(1)}\\
 \lambda_{2}^{2K} \phi_1^{(2)}&\lambda_{2}^{2K-1} \phi_2^{(2)}&\dots& \lambda_{2}\phi_2^{(2)}&\phi_1^{(2)}\\
 \vdots&\vdots&\ddots&\vdots&\vdots\\
 \lambda_{2K}^{2K} \phi_1^{(2K)}& \lambda_{2K}^{2K-1} \phi_2^{(2K)}&\dots&
 \lambda_{2K}\phi_2^{(2K)}&\phi_1^{(2K)}\\
 \lambda_{2K+1}^{2K} \phi_1^{(2K+1)}&\lambda_{2K+1}^{2K-1} \phi_2^{(2K+1)}&\dots& \lambda_{2K+1}\phi_2^{(2K+1)}&\phi_1^{(2K+1)}
                \end{array}
              \right)},  \label{TwoK+1sd-DT1} \\
\phi_2[2K+1] &=&\frac{\det\left(
                \begin{array}{ccccc}
 \lambda^{2K+1}\phi_1 &\lambda^{2K} \phi_2&\dots& \lambda \phi_1&\phi_2\\
 \lambda_{1}^{2K+1}\phi_1^{(1)} &\lambda_{1}^{2K} \phi_2^{(1)}&\dots& \lambda_{1}\phi_1^{(1)}&\phi_2^{(1)}\\
 \vdots&\vdots&\ddots&\vdots&\vdots\\
 \lambda_{2K}^{2K+1}\phi_1^{(2K)} &\lambda_{2K}^{2K} \phi_2^{(2K)}&\dots&
 \lambda_{2K}\phi_1^{(2K)}&\phi_2^{(2K)}\\
 \lambda_{2K+1}^{2K+1}\phi_1^{(2K+1)} &\lambda_{2K+1}^{2K} \phi_2^{(2K+1)}&\dots& \lambda_{2K+1}\phi_1^{(2K+1)}&\phi_2^{(2K+1)}
                \end{array}
              \right)
}{\det\left(
                \begin{array}{ccccc}
\lambda_{1}^{2K} \phi_2^{(1)}&\lambda_{1}^{2K-1} \phi_1^{(1)}&\dots& \lambda_{1}\phi_1^{(1)}&\phi_2^{(1)}\\
 \lambda_{2}^{2K} \phi_2^{(2)}&\lambda_{2}^{2K-1} \phi_1^{(2)}&\dots& \lambda_{2}\phi_1^{(2)}&\phi_2^{(2)}\\
 \vdots&\vdots&\ddots&\vdots&\vdots\\
 \lambda_{2K}^{2K} \phi_1^{(2K)}& \lambda_{2K}^{2K-1} \phi_1^{(2K)}&\dots&
 \lambda_{2K}\phi_1^{(2K)}&\phi_2^{(2K)}\\
 \lambda_{2K+1}^{2K} \phi_1^{(2K+1)}&\lambda_{2K+1}^{2K-1} \phi_1^{(2K+1)}&\dots& \lambda_{2K+1}\phi_1^{(2K+1)}&\phi_2^{(2K+1)}
                \end{array}
              \right)}.  \label{TwoK+1sd-DT2}
\end{eqnarray}
Similarly the $(2K+1)$-fold transformed solutions given by
(\ref{TwoK+1sd-DT1})-(\ref{TwoK+1sd-DT2}) satisfy the linear system
(\ref{linear2})
\begin{eqnarray}
\frac{\partial \phi_1[2K+1]}{\partial y} &=&\frac{\partial
\phi_1[2K+1]}{\partial x}+\frac{\mbox{i}}{2}\left(\frac{\partial
s[2K+1]}{\partial
y}-\frac{\partial s[2K+1]}{\partial x}\right)\phi_1[2K+1]+\lambda \phi_2[2K+1],  \nonumber \\
\frac{\partial \phi_2[2K+1]}{\partial y} &=&\frac{\partial
\phi_2[2K+1]}{\partial x}+\lambda
\phi_1[2K+1]-\frac{\mbox{i}}{2}\left(\frac{\partial
s[2K+1]}{\partial
y}-\frac{\partial s[2K+1]}{\partial x}\right)\phi_2[2K+1],  \label{TwosdK+1-Lax1 DT} \\
\frac{\partial \phi_1[2K+1]}{\partial t} &=&\frac{\partial
\phi_1[2K+1]}{\partial
x}+ \frac{1}{4\lambda}e^{\mbox{i}s[2K+1]}\phi_2[2K+1],  \nonumber \\
\frac{\partial \phi_2[2K+1]}{\partial t} &=&\frac{\partial
\phi_2[2K+1]}{\partial
x}+\frac{1}{4\lambda}e^{-\mbox{i}s[2K+1]}\phi_1[2K+1].
\label{TwosdK+1-Lax2 DT}
\end{eqnarray}%
Using equations (\ref{TwoK+1sd-DT1})-(\ref{TwoK+1sd-DT2}) in
(\ref{TwosdK+1-Lax1 DT})-(\ref{TwosdK+1-Lax2 DT}), we obtain
\begin{equation}
s\left[2K+1\right]=s+2\mbox{i}\ln\frac{\det\left(
                \begin{array}{cccccc}
 \lambda_{1}^{2K}\phi_2^{(1)} &\lambda_{1}^{2K-1} \phi_1^{(1)}&\dots& \lambda_{1}\phi_1^{(1)}&\phi_2^{(1)}\\
 \lambda_{2}^{2K}\phi_2^{(2)} &\lambda_{2}^{2K-1} \phi_1^{(2)}&\dots& \lambda_{2}\phi_1^{(2)}&\phi_2^{(2)}\\
 \vdots&\vdots&\ddots&\vdots&\vdots\\
 \lambda_{2K}^{2K}\phi_2^{(2K)} &\lambda_{2K}^{2K-1} \phi_1^{(2K)}&\dots&
 \lambda_{2K}\phi_1^{(2K)}&\phi_2^{(2K)}\\
 \lambda_{2K+1}^{2K}\phi_2^{(2K+1)} &\lambda_{2K+1}^{2K-1} \phi_1^{(2K+1)}&\dots& \lambda_{2K+1}\phi_1^{(2K+1)}&\phi_2^{(2K+1)}
                \end{array}
              \right)}{\det\left(
                \begin{array}{cccccc}
 \lambda_{1}^{2K}\phi_1^{(1)} &\lambda_{1}^{2K-1} \phi_2^{(1)}&\dots& \lambda_{1}\phi_2^{(1)}&\phi_1^{(1)}\\
 \lambda_{2}^{2K}\phi_1^{(2)} &\lambda_{2}^{2K-1} \phi_2^{(2)}&\dots& \lambda_{2}\phi_2^{(2)}&\phi_1^{(2)}\\
 \vdots&\vdots&\ddots&\vdots&\vdots\\
 \lambda_{2K}^{2K}\phi_1^{(2K)} &\lambda_{2K}^{2K-1} \phi_2^{(2K)}&\dots&
 \lambda_{2K}\phi_2^{(2K)}&\phi_1^{(2K)}\\
 \lambda_{2K+1}^{2K}\phi_1^{(2K+1)} &\lambda_{2K+1}^{2K-1} \phi_2^{(2K+1)}&\dots& \lambda_{2K+1}\phi_2^{(2K+1)}&\phi_1^{(2K+1)}
                \end{array}
              \right)}. \label{2K+1fold-sg16}
\end{equation}%
In the following section, we shall derive explicit expressions of
first two nontrivial solutions and demonstrate our results
graphically for different choices of parameters and arbitrary
functions.

\section{Explicit solutions}
In order to compute higher-order nontrivial solutions of the $2$D
SGE we should start from a trivial solution, i.e., $s=0$. Under this
choice the linear system (\ref{linear2}) becomes
\begin{eqnarray}
\frac{\partial }{\partial y}\left( \begin{array}{c} \phi_1 \\
\phi_2 \end{array}\right)&=&\frac{\partial}{\partial
x}\left( \begin{array}{c} \phi_1 \\
\phi_2 \end{array}\right)+ \left(\begin{array}{cc} 0 & \lambda
\\
\lambda & 0 \end {array}\right) \left( \begin{array}{c} \phi_1 \\
\phi_2 \end{array}\right),\nonumber \\
\frac{\partial }{\partial t}\left( \begin{array}{c} \phi_1 \\
\phi_2 \end{array}\right)&=&\frac{\partial}{\partial
x}\left( \begin{array}{c} \phi_1 \\
\phi_2 \end{array}\right)+ \left(\begin{array}{cc} 0 &
\frac{1}{4\lambda}
\\
\frac{1}{4\lambda} & 0 \end {array}\right) \left( \begin{array}{c} \phi_1 \\
\phi_2 \end{array}\right). \label{solution}
\end{eqnarray}
Integration of above linear system yields
\begin{equation}
\phi_1(x,y,t;\lambda)= Ae^{\chi}+Be^{-\chi}, \quad \quad \quad \quad
\quad \phi_2(x,y,t;\lambda)=Ae^{\chi}-Be^{-\chi},
\end{equation}
where $\chi=\lambda y+\frac{1}{4\lambda}t+g(x,y,t)$ and $g(x,y,t)$
is a function of the form $g(x,y,t)=f\left(x+y+t\right)$.

\subsection{First-order solutions}
\begin{figure}
\centering
\includegraphics{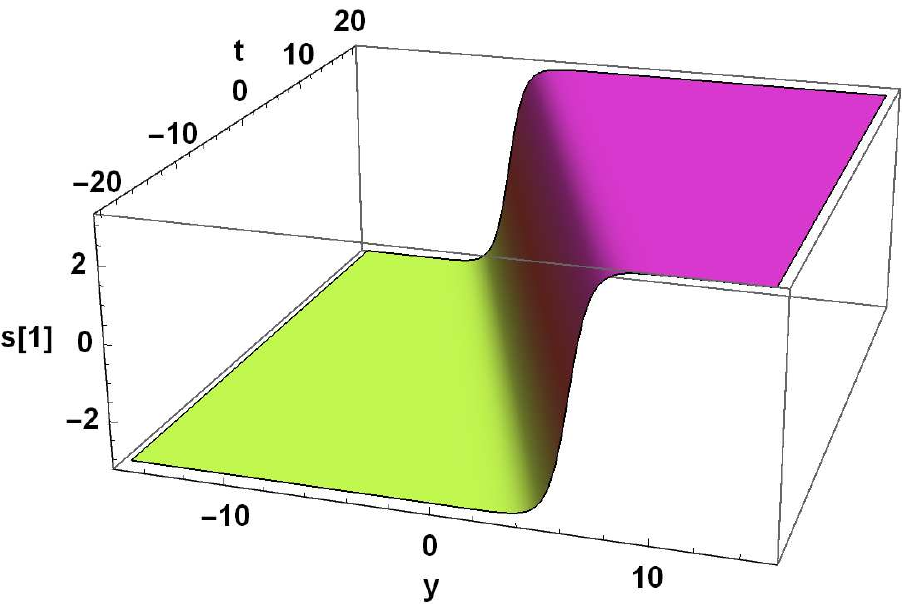}
   \caption{One-kink solution (\ref{onefoldEXP}) of $2$D SGE (\ref{SGE1}) for $\lambda_1=0.9$ and $f(x+y+t)=x+y+t$ at $x=0$.}\label{kink1}
\end{figure}

\begin{figure}
\centering
\includegraphics{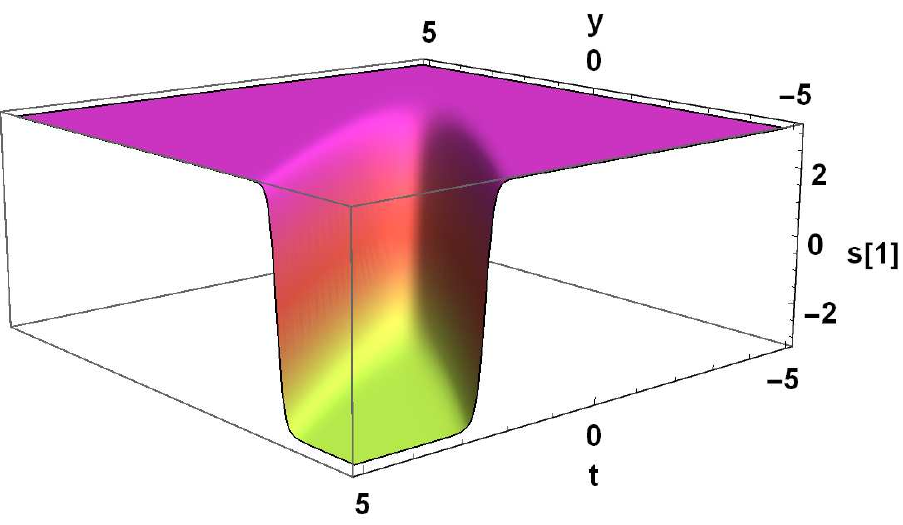}
   \caption{One-kink solution (\ref{onefoldEXP}) of $2$D SGE (\ref{SGE1}) for $\lambda_1=1.5$ and $f(x+y+t)=(x+y+t)^2$ at $x=0$.}\label{kink2}
\end{figure}

\begin{figure}
\centering
\includegraphics{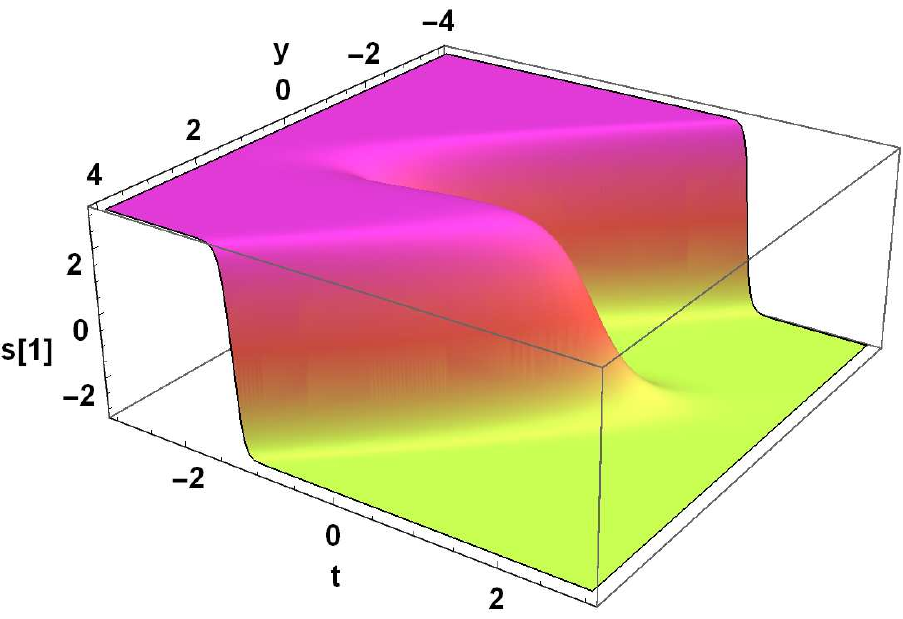}
   \caption{One-kink solution (\ref{onefoldEXP}) of $2$D SGE (\ref{SGE1}) for $\lambda_1=2$ and $f(x+y+t)=\frac{x+y+t-(x+y+t)^3}{2}$ at $x=0$.}\label{kinkSG}
\end{figure}

\begin{figure}
\centering
\includegraphics{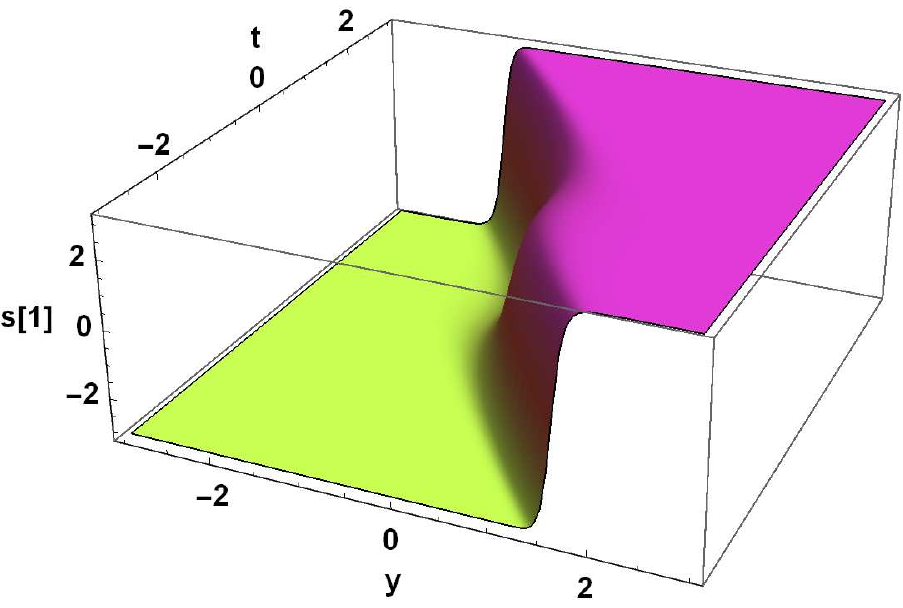}
   \caption{One-kink solution (\ref{onefoldEXP}) of $2$D SGE (\ref{SGE1}) for $\lambda_1=2$ and $f(x+y+t)=(x+y+t)^3$ at $x=0$.}\label{kinkS}
\end{figure}

\begin{figure}
\centering
\includegraphics{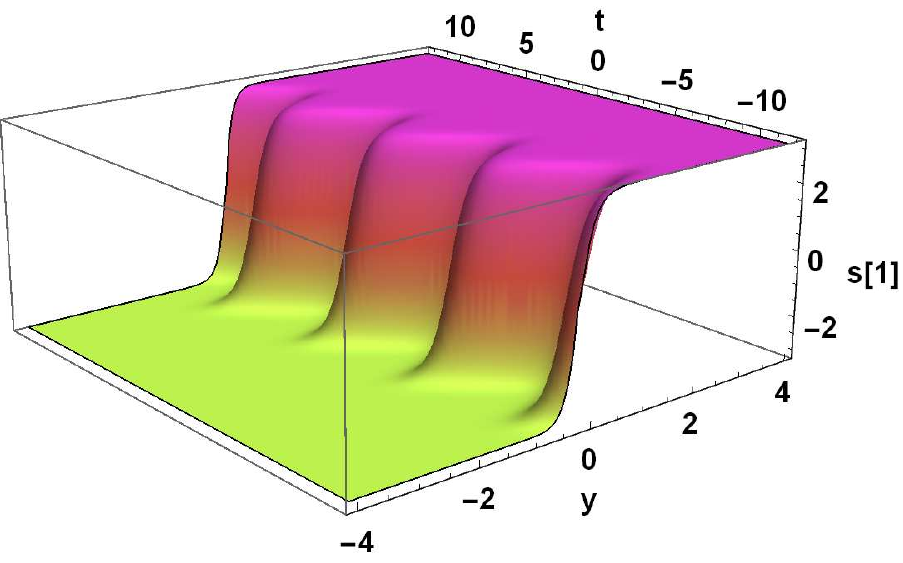}
   \caption{One-kink solution (\ref{onefoldEXP}) of $2$D SGE (\ref{SGE1}) for $\lambda_1=3$ and $f(x+y+t)=\sin (x+y+t)+\cos (x+y+t)$ at $x=0$.}\label{kinkPeriodic}
\end{figure}

Substituting $\phi_1^{(1)}$ and $\phi_2^{(1)}$, the particular
solutions of the linear system at $\lambda_1$ in equation
(\ref{sg16}), we get first-order nontrivial solution of $2$D SGE
given by
\begin{equation}
s[1]=-2\mbox{i} \ln \left(\frac{A_1e^{\lambda_1
y+\frac{1}{4\lambda_1}t+f(x+y+t)}-B_1 e^{-\lambda_1
y-\frac{1}{4\lambda_1}t-f(x+y+t)}}{A_1e^{\lambda_1
y+\frac{1}{4\lambda_1}t+f(x+y+t)}+B_1 e^{-\lambda_1
y-\frac{1}{4\lambda_1}t-f(x+y+t)}}\right),
\end{equation}
or
\begin{equation}
s[1]=-2\mbox{i} \ln \left(\frac{A_1e^{2\lambda_1
y+\frac{1}{2\lambda_1}t+2f}-B_1}{A_1e^{2\lambda_1
y+\frac{1}{2\lambda_1}t+2f}+B_1}\right), \label{onefoldEXP}
\end{equation}
is the required explicit expression of first-order nontrivial
solution of $2$D SGE. Here $f(x,y,t)$ is an arbitrarily chosen
function that satisfies equation (\ref{solution}). Profiles of
one-kink soliton solution of $2$D SGE (\ref{SGE1}) are shown in
figures (\ref{kink1})-(\ref{kinkPeriodic}) for $A_1=-B_2=1$ and
$A_2=B_1=\mbox{i}$. Fig. (\ref{kink1}) represents a kink soliton
plotted for $\lambda_1=0.9$. Fig. (\ref{kink2}) indicates a
propagation of one-kink on a parabolic path for $\lambda_1=1.5$.
Figs. (\ref{kinkSG}) illustrates s-shaped kinks for $\lambda_1=2$.
Similarly Fig. \ref{kinkS} represents bending of one-kink from its
path during propagation in nonlinear medium whereas Fig.
\ref{kinkPeriodic} display oscillating one-kink for $\lambda_1=3$.
(All choices of arbitrary function mentioned in Figures captions)

\subsection{Second-order nontrivial solutions}
Substituting $\phi_1^{(1)}=A_1e^{\theta_1}+B_1 e^{-\theta_1}$,
$\phi_2^{(1)}=A_2e^{\theta_1}-B_2 e^{-\theta_1}$,
$\phi_1^{(2)}=A_3e^{\theta_2}+B_3 e^{-\theta_2}$,
$\phi_2^{(2)}=A_4e^{\theta_2}-B_4 e^{-\theta_2}$,
$A_1=-A_3=-B_2=B_4=1$ and $A_2=A^{\ast}_4=B_1=B^{\ast}_3=\mbox{i}$,
in equation (\ref{Twosg16}), we have,
\begin{equation}
s[2]=4\tan^{-1}
\left(\left(\frac{\lambda_1+\lambda_2}{\lambda_1-\lambda_2}\right)\frac{\sinh
\left(\theta_1-\theta_2\right)}{\cosh
\left(\theta_1+\theta_2\right)}\right), \label{2foldEXP}
\end{equation}
where $\theta_1=\lambda_1y+\frac{1}{4\lambda_1}t+f_{1}(x+y+t)$ and
$\theta_2=\lambda_2y+\frac{1}{4\lambda_2}t+f_{2}(x+y+t)$.

\begin{figure}
\centering
\includegraphics{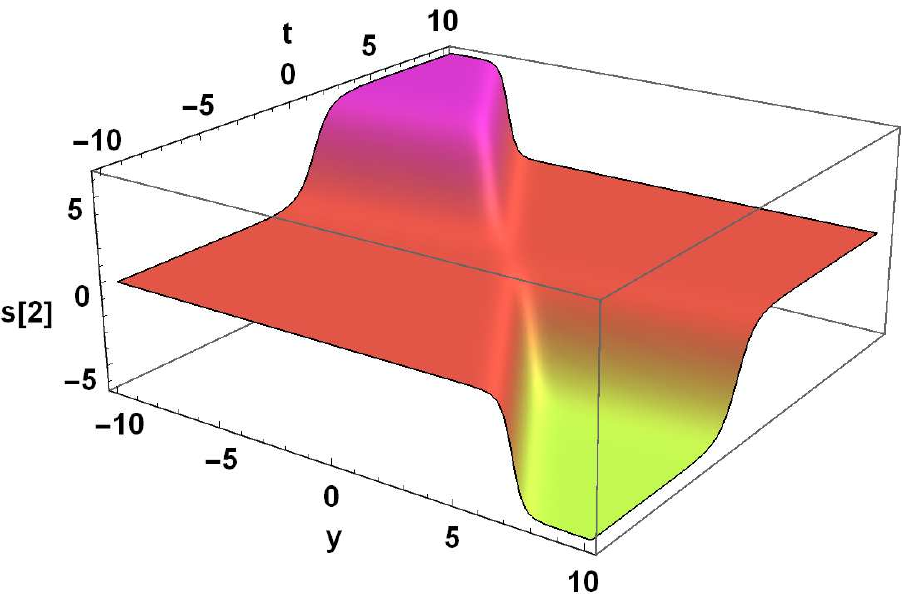}
   \caption{Attractive type of interaction of kinks of $2$D SGE (\ref{SGE1}) for $f_{1}=f_{2}=x+y+t$ at $x=0$.}\label{RKKBL}
\end{figure}

\begin{figure}
\centering
\includegraphics{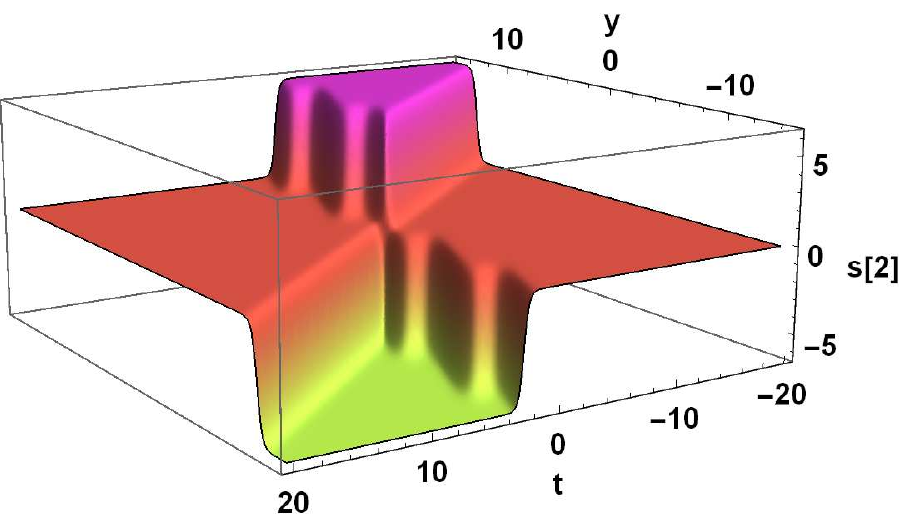}
   \caption{Attractive  type of interaction of kinks of $2$D SGE (\ref{SGE1}) for $f_{1}= \left(x+y+t\right)$ and $f_{2}=\sin \left(x+y+t\right)$ at $x=0$.}\label{RKKSC}
\end{figure}

\begin{figure}
\centering
\includegraphics{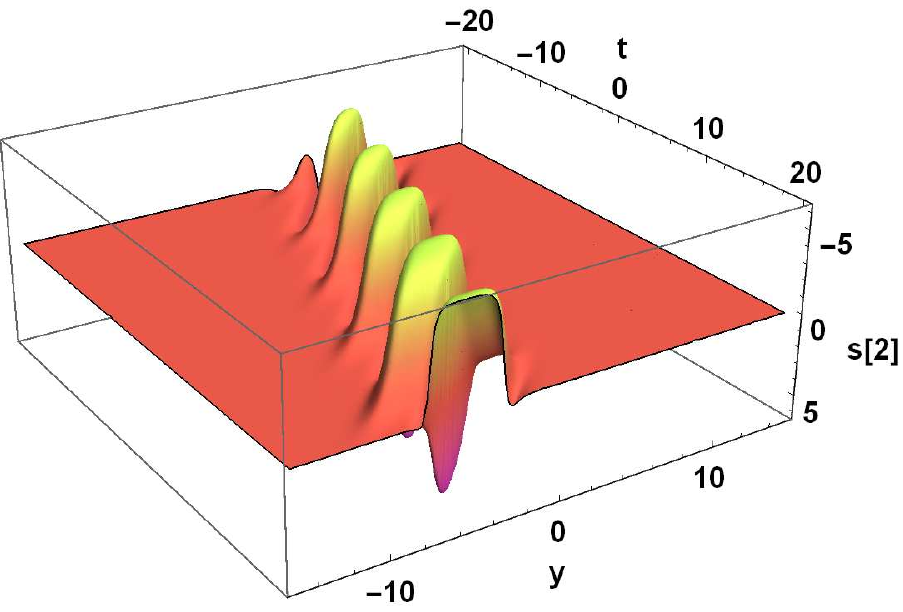}
   \caption{Attractive type of interaction of kinks of $2$D SGE (\ref{SGE1}) for $f_{1}=\cos \left(x+y+t\right)$ and $f_{2}=\sin \left(x+y+t\right)$ at $x=0$.}\label{AKAKSC}
\end{figure}

\begin{figure}
\centering
\includegraphics{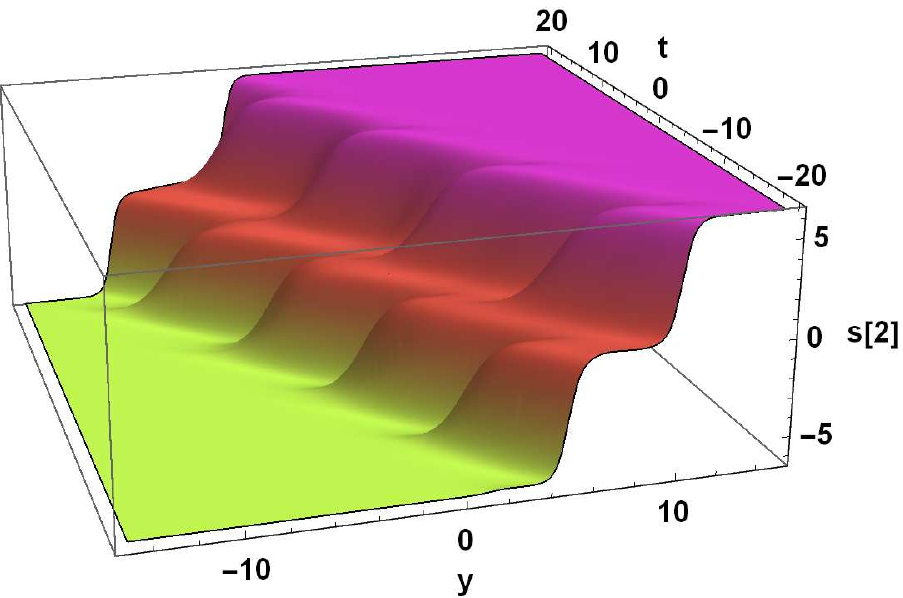}
   \caption{Repulsive type of interaction of kinks of $2$D SGE (\ref{SGE1}) for $f_{1}=\cos\left(x+y+t\right)$ and $f_{2}=\sin \left(x+y+t\right)$ at $x=0$.}\label{AKAKLSin}
\end{figure}

Kinks display two types of interactions, namely, repulsive and
attractive. Interaction of kinks are seeking great attention due to
its wide implications in many branches of science such as particle
physics and biological sciences. Two-kink solutions usually appear
in the problems whose respective potential is controlled by some
parameters instead of fields. In kink-kink interactions when two
kinks approaches each other, they exert repulsive force on each
other and reflect back with an equal but opposite velocities to
their initial velocities. Since kink soliton exhibits particle-like
characteristics in this way interaction of kink and anti-kink can be
perceived as an interaction between particle and anti-particle. In
attractive type of interactions, kink and anti-kink attempt to
annihilate when they get closer. Interactions of kinks for various
choices of arbitrary functions are exhibited in Figs. (\ref{RKKBL}),
(\ref{RKKSC}) and (\ref{AKAKSC}). Fig. (\ref{RKKBL}) represents an
attractive type of interaction of two kinks for $\lambda_1=1$ and
$\lambda_2=-1.09$.  Similarly, Figs. (\ref{RKKSC}) and
(\ref{AKAKSC}) also demonstrate attractive type of interactions
obtained for the parameters $\lambda_1=1, \lambda_2=0.2$ and
$\lambda_1=1, \lambda_2=0.9$, respectively. Where as Fig.
(\ref{AKAKLSin}) displays a repulsive type of interaction of two
kinks for parameters $\lambda_1=0.9$ and $\lambda_2=-0.8$.

\begin{figure}
\centering
\includegraphics{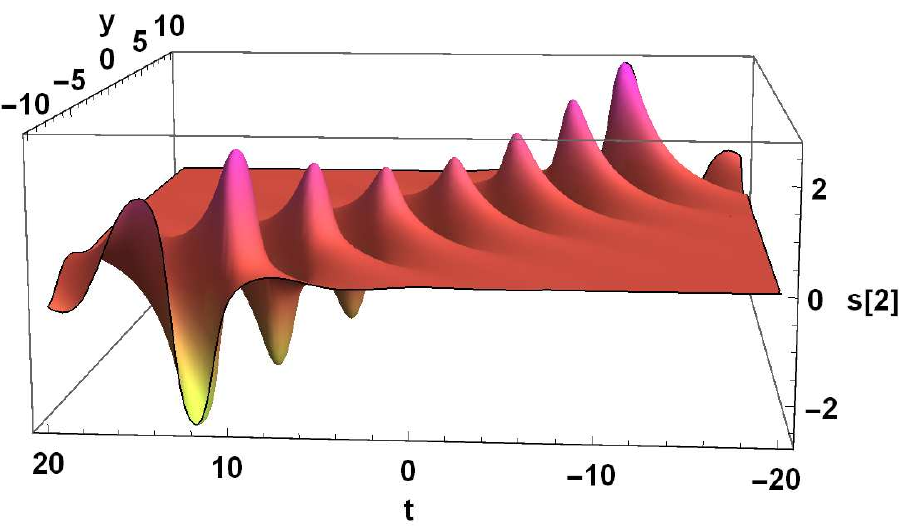}
   \caption{Breather solution of $2$D SGE (\ref{SGE1}) at $f_{1}=0$ and $f_{2}=x+y+t$.}\label{BreatherL0}
\end{figure}

\begin{figure}
\centering
\includegraphics{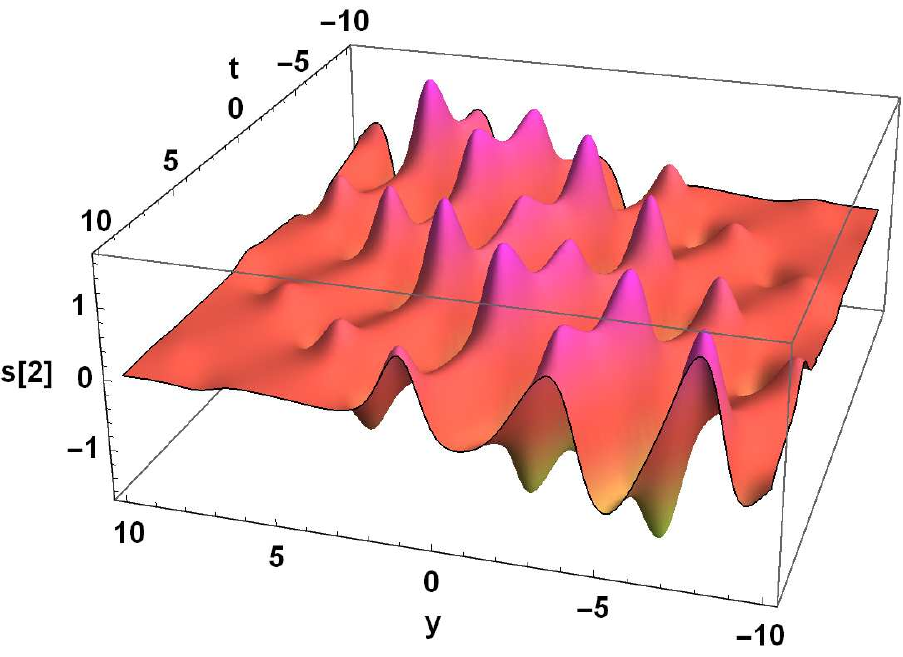}
   \caption{Breather solution of $2$D SGE (\ref{SGE1}) at $f_{1}=0$ and $f_{2}=\sin (x+y+t)$.}\label{Breather0Sin}
\end{figure}

\begin{figure}
\centering
\includegraphics{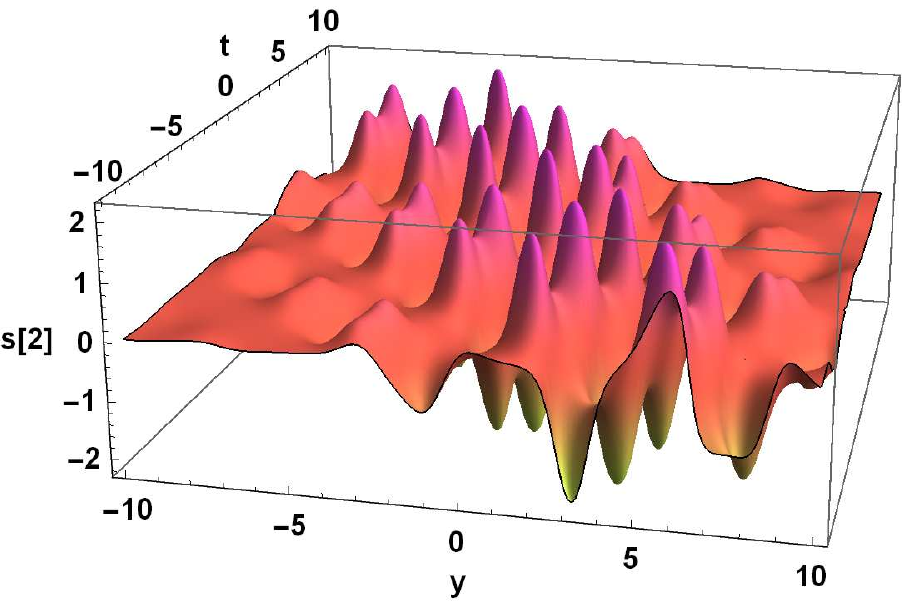}
   \caption{Breather solution of $2$D SGE (\ref{SGE1}) at $f_{1}=\sin (x+y+t)$ and $f_{2}=\cos (x+y+t)$.}\label{BreatherSC}
\end{figure}

When two kinks fused together in such a way that they may not
preserve their shape, consequently a bound state is formed, which is
known as breather solutions. An explicit expression of first-order
breather is obtained by taking $\lambda_2=\lambda^{\ast}_1$ in
expression (\ref{2foldEXP}), i.e.,
\begin{equation}
s[2]=4\tan^{-1} \left(\left(\frac{\mbox{Re}
(\lambda_1)}{\mbox{i}\mbox{Im}(\lambda_1)}\right)\frac{\sinh
\left(\theta_1-\theta^{\ast}_1\right)}{\cosh
\left(\theta_1+\theta^{\ast}_1\right)}\right). \label{BreatherEXP}
\end{equation}
Different profiles of (\ref{BreatherEXP}) are represented in Figs.
(\ref{BreatherL0})-(\ref{BreatherSC}) for spectral parameter
$\lambda_1=0.2+0.6\mbox{i}$, $\lambda_1=0.2+0.7\mbox{i}$,
$\lambda_1=0.2+0.7\mbox{i}$ and $\lambda_1=0.5+0.5\mbox{i}$
respectively.

\section{Conclusions}
In this article, we have derived a general formula for $N$-soliton
solution of sine-Gordon equation (\ref{SGE1}) in $2+1$-dimensions
from sequential application of Darboux transformation on associated
linear eigenvalue problem. Explicit expressions of one- and
two-soliton solutions have computed for the our model. In order to
illustrate our results we have presented dynamics of single and
different profile two-soliton interactions for different choices of
spectral parameters as well as for the arbitrary functions. We are
expecting that solutions obtained in this article will be helpful in
explaining various physical phenomena in higher dimensions, for
example, the dynamics of DNA, dynamics of crystalline latices near
dislocation etc. There are several interesting directions, for
example, one can explore the multi-soliton solutions of some
well-known integrable equations such as nonlinear Schr\"{o}dinger
equation, the KdV equation, short pulse equation in higher
dimensions. Similarly one can also calculate higher order degenerate
solutions of sine-Gordon equation (\ref{SGE1}) in $2+1$-dimensions.
We will address these open problems in near future.

\bibliographystyle{IEEetran}

\bibliography{bibSGE7}

\begin{thebibliography}{10}
\providecommand{\url}[1]{#1}
\csname url@samestyle\endcsname
\providecommand{\newblock}{\relax}
\providecommand{\bibinfo}[2]{#2}
\providecommand{\BIBentrySTDinterwordspacing}{\spaceskip=0pt\relax}
\providecommand{\BIBentryALTinterwordstretchfactor}{4}
\providecommand{\BIBentryALTinterwordspacing}{\spaceskip=\fontdimen2\font plus
\BIBentryALTinterwordstretchfactor\fontdimen3\font minus
  \fontdimen4\font\relax}
\providecommand{\BIBforeignlanguage}[2]{{%
\expandafter\ifx\csname l@#1\endcsname\relax
\typeout{** WARNING: IEEEtran.bst: No hyphenation pattern has been}%
\typeout{** loaded for the language `#1'. Using the pattern for}%
\typeout{** the default language instead.}%
\else
\language=\csname l@#1\endcsname
\fi
#2}}
\providecommand{\BIBdecl}{\relax}
\BIBdecl

\bibitem{Rajaraman}
R.~Rajaraman, ``Solitons and instantons,'' 1982.

\bibitem{Bour1862}
E.~Bour, \emph{Th{\'e}orie de la d{\'e}formation des surfaces}.\hskip 1em plus
  0.5em minus 0.4em\relax Gauthier-Villars, 1891.

\bibitem{Victor}
A.~B{\"a}cklund, ``{\"U}ber fl{\"a}chentransformationen math,'' \emph{Ann},
  vol.~9, pp. 297--320, 1876.

\bibitem{Frenkel}
J.~Frenkel and T.~Kontorova, ``On the theory of plastic deformation and
  twinning,'' \emph{Izv. Akad. Nauk, Ser. Fiz.}, vol.~1, pp. 137--149, 1939.

\bibitem{Ablowitz1973}
M.~J. Ablowitz, D.~J. Kaup, A.~C. Newell, and H.~Segur, ``Nonlinear-evolution
  equations of physical significance,'' \emph{Physical Review Letters},
  vol.~31, no.~2, p. 125, 1973.

\bibitem{Zakharov}
V.~E. Zakharov, L.~A. Takhtadzhyan, and L.~D. Faddeev, ``Complete description
  of solutions of the "sine-gordon" equation,'' in \emph{Doklady Akademii
  Nauk}, vol. 219, no.~6.\hskip 1em plus 0.5em minus 0.4em\relax Russian
  Academy of Sciences, 1974, pp. 1334--1337.

\bibitem{Lamb1967}
G.~Lamb~Jr, ``Propagation of ultrashort optical pulses,'' \emph{Physics Letters
  A}, vol.~25, no.~3, pp. 181--182, 1967.

\bibitem{CMP1}
B.~D. Josephson, ``Supercurrents through barriers,'' \emph{Advances in
  Physics}, vol.~14, no.~56, pp. 419--451, 1965.

\bibitem{CMP2}
P.~Lebwohl and M.~Stephen, ``Properties of vortex lines in superconducting
  barriers,'' \emph{Physical Review}, vol. 163, no.~2, p. 376, 1967.

\bibitem{CMP3}
A.~Scott, ``Steady propagation on long josephson junctions,'' \emph{Bull. Am.
  Phys. Soc}, vol.~12, p. 308, 1967.

\bibitem{CMP4}
A.~Scott and W.~Johnson, ``Internal flux motion in large josephson junctions,''
  \emph{Applied Physics Letters}, vol.~14, no.~10, pp. 316--318, 1969.

\bibitem{CMP5}
A.~Barone, ``Flux-flow effect in josephson tunnel junctions,'' \emph{Journal of
  Applied Physics}, vol.~42, no.~7, pp. 2747--2751, 1971.

\bibitem{particle1}
T.~Skyrme, ``A non-linear theory of strong interactions,'' \emph{Proceedings of
  the Royal Society of London. Series A. Mathematical and Physical Sciences},
  vol. 247, no. 1249, pp. 260--278, 1958.

\bibitem{particle1a}
------, ``Particle states of a quantized meson field,'' \emph{Proceedings of
  the Royal Society of London. Series A. Mathematical and Physical Sciences},
  vol. 262, no. 1309, pp. 237--245, 1961.

\bibitem{particle2}
J.~Rubinstein, ``Sine-gordon equation,'' \emph{Journal of Mathematical
  Physics}, vol.~11, no.~1, pp. 258--266, 1970.

\bibitem{particle3}
U.~Enz, ``Discrete mass, elementary length, and a topological invariant as a
  consequence of a relativistic invariant variational principle,''
  \emph{Physical Review}, vol. 131, no.~3, p. 1392, 1963.

\bibitem{particle4}
N.~Rosen and H.~B. Rosenstock, ``The force between particles in a nonlinear
  field theory,'' \emph{Physical Review}, vol.~85, no.~2, p. 257, 1952.

\bibitem{BIO1}
V.~G. Ivancevic, T.~T. Ivancevic \emph{et~al.}, ``Sine--gordon solitons, kinks
  and breathers as physical models of nonlinear excitations in living cellular
  structures,'' \emph{Journal of Geometry and Symmetry in Physics}, vol.~31,
  pp. 1--56, 2013.

\bibitem{BIO2}
S.~Yomosa, ``Soliton excitations in deoxyribonucleic acid (dna) double
  helices,'' \emph{Physical Review A}, vol.~27, no.~4, p. 2120, 1983.

\bibitem{BIO3}
M.~Peyrard and A.~R. Bishop, ``Statistical mechanics of a nonlinear model for
  dna denaturation,'' \emph{Physical review letters}, vol.~62, no.~23, p. 2755,
  1989.

\bibitem{BIO4}
L.~V. Yakushevich, \emph{Nonlinear physics of DNA}.\hskip 1em plus 0.5em minus
  0.4em\relax John Wiley \& Sons, 2006.

\bibitem{BIO5}
M.~Daniel and V.~Vasumathi, ``Solitonlike base pair opening in a helicoidal
  dna: An analogy with a helimagnet and a cholesteric liquid crystal,''
  \emph{Physical Review E}, vol.~79, no.~1, p. 012901, 2009.

\bibitem{Gibbon}
J.~Gibbon, I.~James, and I.~M. Moroz, ``An example of soliton behaviour in a
  rotating baroclinic fluid,'' \emph{Proceedings of the Royal Society of
  London. A. Mathematical and Physical Sciences}, vol. 367, no. 1729, pp.
  219--237, 1979.

\bibitem{book1}
V.~B. Matveev and V.~Matveev, ``Darboux transformations and solitons,'' 1991.

\bibitem{book2}
C.~Rogers, C.~Rogers, and W.~Schief, \emph{B{\"a}cklund and Darboux
  transformations: geometry and modern applications in soliton theory}.\hskip
  1em plus 0.5em minus 0.4em\relax Cambridge University Press, 2002, no.~30.

\bibitem{book3}
C.~Gu, H.~Hu, A.~Hu, and Z.~Zhou, \emph{Darboux transformations in integrable
  systems: theory and their applications to geometry}.\hskip 1em plus 0.5em
  minus 0.4em\relax Springer Science \& Business Media, 2004.

\bibitem{Novikov1984}
A.~Veselov and S.~Novikov, ``Finite-zone, two-dimensional, potential
  schr{\"o}dinger operators. explicit formulas and evolution equations,'' in
  \emph{Soviet Math. Dokl}, vol.~30, 1984, pp. 588--591.

\bibitem{Stewartson}
A.~Davey and K.~Stewartson, ``On three-dimensional packets of surface waves,''
  \emph{Proceedings of the Royal Society of London. A. Mathematical and
  Physical Sciences}, vol. 338, no. 1613, pp. 101--110, 1974.

\bibitem{2DSGE1}
G.~Wang, K.~Yang, H.~Gu, F.~Guan, and A.~Kara, ``A (2+ 1)-dimensional
  sine-gordon and sinh-gordon equations with symmetries and kink wave
  solutions,'' \emph{Nuclear Physics B}, vol. 953, p. 114956, 2020.

\bibitem{2DSGE2}
A.-M. Wazwaz, ``New integrable (2+ 1)-dimensional sine-gordon equations with
  constant and time-dependent coefficients: Multiple optical kink wave
  solutions,'' \emph{Optik}, vol. 216, p. 164640, 2020.

\bibitem{2DSGE3}
------, ``New integrable (2+ 1)-and (3+ 1)-dimensional sinh-gordon equations
  with constant and time-dependent coefficients,'' \emph{Physics Letters A},
  vol. 384, no.~23, p. 126529, 2020.

\bibitem{2DSGE4}
Y.~Feng, S.~Bilige, and X.~Wang, ``Diverse exact analytical solutions and novel
  interaction solutions for the (2+ 1)-dimensional ito equation,''
  \emph{Physica Scripta}, vol.~95, no.~9, p. 095201, 2020.

\bibitem{2DSGE5}
G.~Wang and A.~Kara, ``A (2+ 1)-dimensional kdv equation and mkdv equation:
  symmetries, group invariant solutions and conservation laws,'' \emph{Physics
  Letters A}, vol. 383, no.~8, pp. 728--731, 2019.

\end{thebibliography}
\end{document}